\documentclass[twocolumn]{aa}
\usepackage{graphicx}
\usepackage[]{natbib}
\newcommand{\Prel}{P_\mathrm{rel}}
\newcommand\pcc{cm$^{-3}$}
\newcommand{\Tetwo}{T_{\mathrm{e2}}}
\newcommand{\Tptwo}{T_{\mathrm{p2}}}
\newcommand\too{\, \rightarrow \, }
\newcommand\kmps{km s$^{-1}$}
\newcommand\Rsk{R_{\mathrm{sk}}}
\newcommand\fRad{f_\mathrm{sk}}
\newcommand\tSNR{t_{\mathrm{snr}}}
\newcommand\rg{r_{g}}
\newcommand\etamfp{\eta_\mathrm{mfp}}
\newcommand\muG{$\mu$G}
\newcommand\alf{Alfv\'en}
\newcommand\pmax{p_\mathrm{max}}
\newcommand\Emax{E_{\mathrm{max}}}
\newcommand\Vsk{u_0}
\newcommand\uUP{u_\mathrm{up}}
\newcommand\uDS{u_\mathrm{ds}}
\newcommand\Rtot{r_\mathrm{tot}}
\newcommand\Rsub{r_\mathrm{sub}}
\newcommand\EnSN{E_{\mathrm{sn}}}
\newcommand\EnCR{E_{\mathrm{cr}}}
\newcommand\fHe{f_\mathrm{He}}
\newcommand\Mej{M_{\mathrm{ej}}}
\newcommand\Msun{M_{\odot}}
\newcommand\tstart{t_{\mathrm{start}}}
\newcommand\tStart{t_{\mathrm{start}}}
\newcommand{\rel}{relativistic}

\newcommand\NL{nonlinear}
\newcommand{\npz}{n_\mathrm{p0}}
\newcommand{\nHe}{n_\mathrm{He}}
\newcommand{\Ndt}{N(p,\Delta t)}
\newcommand{\tch}{t_\mathrm{ch}}
\newcommand{\Rch}{R_\mathrm{ch}}
\newcommand{\Vch}{V_\mathrm{ch}}

\newcommand{\facTh}{f_{\mathrm{th}}}
\newcommand{\rhoUP}{\rho_{\mathrm{up}}}
\newcommand{\rhoDW}{\rho_{\mathrm{dw}}}
\newcommand{\dsa}{diffusive shock acceleration} 
 
\newcommand{\rhoISM}{\rho_{\mathrm{ISM}}}
\newcommand{\radFS}{R_{\mathrm{FS}}}

\newcommand\syn{synchrotron}
\newcommand\synch{synchrotron}
\newcommand\brem{bremsstrahlung}

\newcommand\IC{inverse-Compton}
\newcommand\pion{pion-decay}
\newcommand\GGG{G347.3$-$0.5}
\newcommand\iec{i.e.,}
\newcommand\egc{e.g.,}
\newcommand\etal{et al.}
\newcommand{\xx}[1]{\times 10^{#1}}
\newcommand\gameff{\gamma_\mathrm{eff}}

\newcommand\etaTP{\eta_\mathrm{TP}}
\newcommand\etainj{\eta_\mathrm{inj}}
\newcommand{\EnDen}{\rho_\mathrm{en}}

\newcount\listnorom
\newcommand\listromanDE{\global\advance \listnorom by 1 
{\lowercase\expandafter{(\romannumeral\listnorom)}\ }}
\newcommand\newlistroman{\listnorom=0}

\begin{document}
   \title{Hydrodynamic Simulation of Supernova
Remnants Including Efficient Particle Acceleration}

   \author{Donald C.~Ellison,
          \inst{1}
 Anne Decourchelle\inst{2}
          \and
          Jean Ballet\inst{2} }

   \offprints{D. Ellison}

   \institute{Department of Physics, North Carolina State
 University, Box 8202, Raleigh NC 27695, U.S.A.\\
              \email{don\_ellison@ncsu.edu}
         \and
Service d'Astrophysique, DSM/DAPNIA, CEA Saclay,
91191 Gif-sur-Yvette, France\\
             \email{adecourchelle@cea.fr, jballet@cea.fr} }

   \date{Received July 11, 2003; accepted August 14, 2003}

   \abstract{
A number of supernova remnants (SNRs) show nonthermal X-rays assumed
to be synchrotron emission from shock accelerated TeV electrons.  The
existence of these TeV electrons strongly suggests that the shocks in
SNRs are sources of galactic cosmic rays (CRs).
In addition, there is convincing evidence from broad-band studies of
individual SNRs and elsewhere that the particle acceleration process
in SNRs can be efficient and nonlinear.  If SNR shocks are efficient
particle accelerators, the production of CRs impacts the thermal
properties of the shock heated, X-ray emitting gas and the SNR
evolution.
We report on a technique that couples nonlinear diffusive shock
acceleration, including the backreaction of the accelerated particles
on the structure of the forward and reverse shocks, with a
hydrodynamic simulation of SNR evolution.
Compared to models which ignore CRs, the most important hydrodynamical
effects of placing a significant fraction of shock energy into CRs are
larger shock compression ratios and lower temperatures in the shocked
gas.
We compare our results, which use an approximate description of the
acceleration process, with a more complete model where the full CR
transport equations are solved \citep[i.e.,][]{BKV2002}, and find
excellent agreement for the CR spectrum summed over the SNR lifetime
and the evolving shock compression ratio.  The importance of the
coupling between particle acceleration and SNR dynamics for the
interpretation of broad-band continuum and thermal X-ray observations
is discussed.
   \keywords{ISM: cosmic rays --- acceleration of
particles --- shock waves --- ISM: supernova remnants --- X-rays: ISM}
   }

\titlerunning{Efficient particle acceleration in SNRs}

   \maketitle
%
%________________________________________________________________

\section{Introduction}
It is commonly believed that the shocks in supernova remnants (SNRs)
produce the majority of galactic cosmic rays (CRs) with energies below
$\sim 10^{15}$ eV via diffusive shock acceleration
\citep[e.g.,][]{Drury83,BE87}.
Convincing support for the production of TeV electrons in SNRs comes
from the \syn\ interpretation of nonthermal X-ray emission
\citep[\egc][]{Reynolds98}
observed in
an increasing number of young SNRs 
such as SN1006 \citep{Koyama95},
Cas A \citep{Allen97}, \GGG\ \citep{Slane99}, and RCW 86
\citep{Bork2001}.\footnote{As of this writing, there is no
unambiguous evidence for the production of TeV ions in SNRs
\citep[see][for a discussion of SN1006 in this regard]{BKV2002}. The
recent claim that TeV emission from SNR RX J1713.7-3946 (also called
\GGG) observed by CANGAROO II is from pion-decay
\citep[\egc][]{Enomoto2002} is still under debate \citep[see, for
example,][]{RP2002,Butt2002}. }

In addition to their putative role in accelerating cosmic rays, the
shocks in SNRs heat the ambient interstellar medium and ejecta to
X-ray emitting temperatures. 
The interpretation of these X-ray observations leads to inferences for
important quantities such as the supernova (SN) explosion energy,
ejecta mass and composition, ambient densities, shock speed, and rate
of electron and proton equilibration.
That CR production and thermal heating in SNRs may be coupled comes
from the fact that diffusive shock acceleration is intrinsically
efficient in high Mach number shocks if even a small fraction of the
shock heated plasma is injected into the acceleration process
\citep[e.g.,][]{EE84,JE91,BE99,Malkov98,Blasi2002}. 

If strong coupling between particle acceleration and shock heating
occurs, the modeling of efficient particle acceleration in SNRs offers
the possibility of using the high-quality X-ray and $\gamma$-ray data
currently being collected by spacecraft 
(e.g., {\it Chandra}, {\it XMM-Newton}, INTEGRAL) to address
fundamental questions concerning the SNR origin of CRs and the
underlying physics of diffusive shock acceleration, particularly the
injection of thermal particles into the acceleration process.
Furthermore, strong coupling implies that the inferences made from
X-ray observations may differ substantially between interpretations
which include particle acceleration self-consistently and those that
do not.

Despite the expected efficiency of diffusive shock acceleration, X-ray
spectra from SNRs have generally been modeled and interpreted assuming
that the shocks place an {\it in}significant fraction of their energy
in cosmic rays.
Exceptions to this include the early works of \citet{Chev83},
\citet{Heavens84}, and \citet{BC88}.
\citet{Chev83} investigated the effects of cosmic-ray pressure on SNR
dynamics using a two-fluid, self-similar solution with an arbitrary
fraction of thermal gas (adiabatic index $\gamma=5/3$) and
relativistic gas ($\gamma=4/3$).  More recent work has been done by
\citet{DorfiB93} and \citet{Dorfi94}.
In our preliminary work \citep[i.e.,][]{DEB2000}, we developed a model
which coupled the approximate nonlinear (NL) acceleration calculation of
\citet{BE99} with an analytic, self-similar description of the SNR
hydrodynamics \citep[i.e.,][]{Chev83} and a non-equilibrium ionization
calculation of X-ray emission \citep[e.g.,][]{DB94}. We illustrated
this model by fitting ASCA and RXTE observations of Kepler's SNR and
found adequate fits when acceleration was efficient at the forward
shock but inefficient at the reverse shock.

The effects of efficient particle acceleration on SNR hydrodynamics
where calculated in a hydro computer simulation of SNRs by
\citet{BE2001}. This was done by globally changing the effective ratio
of specific heats, $\gameff$, from $5/3$ to values approaching 1, but
did not include coupling between the acceleration and the hydro.  As
$\gameff$ was decreased, the shocked gas became more compressible, the
shock compression ratio increased, and the interaction region between
the forward and reverse shocks narrowed.  \citet{BE2001} were able to
show in two and three-dimensional simulations that if the interaction
region was narrow enough, convective instabilities produced
Rayleigh-Taylor fingers of dense ejecta material which were able to
reach and perturb the forward shock.

Here, we introduce and describe in detail a CR-Hydro model which uses
the same NL acceleration calculation of \citet{BE99}, but replaces the
self-similar description used in \citet{DEB2000} with a 1-D hydro
simulation such as that used by \citet{BE2001}. The simulation is more
general than the analytic approach used by \citet{DEB2000} since it is
not restricted to self-similar evolution and allows for a continuous
change in the acceleration efficiency as the SNR evolves. We show,
however, that when acceleration efficiency is nearly constant during
the self-similar phase, the two models closely correspond, providing
an important check on the validity of both models.
While we do not calculate X-ray thermal spectra in this paper, we
show, with various examples, how the efficient production of CR
protons influences SNR evolution and discuss the implications this has
on the interpretation of X-ray observations.

\section{CR-Hydro Model}
Our CR-Hydro model couples a spherically symmetric hydrodynamic
simulation with a calculation of \NL\ diffusive shock
acceleration. The particle acceleration calculation determines how
energy is divided between the thermal gas and \rel\ particles, and
provides the particle distribution over all energies behind the shock,
as well as the effective ratio of specific heats, $\gameff$, defined
below.
In future work, we will use the electron and ion distribution
functions to calculate the broad-band continuum photon emission from
radio to TeV energies \citep[\egc][]{ESG2001}, and use the
self-consistent thermal properties in a non-equilibrium calculation of
X-ray lines \citep[\egc][]{DEB2000}.  For now, we restrict ourselves
to calculating CR proton spectra integrated over the SNR
lifetime.

\subsection{Hydrodynamic Simulation}
We use a standard hydrodynamic simulation in one dimension to model
the effects of a supernova explosion in the interstellar medium (ISM)
\citep[see][and references therein]{BE2001}.  We are free to choose
arbitrary ejecta and ISM mass density profiles, but to facilitate our
comparisons discussed below, we adopt the parameters determined for
SN1006 by \citet{BKV2002}, i.e., we assume a constant density,
constant temperature ISM, and take the initial ejecta density profile
to be $\rho \propto r^{-n}$, with $n=7$ and a constant density plateau
at small radii \citep[i.e.,][]{Chev82a}.\footnote{At the start of the
simulation, the ejecta temperature is assumed to be low enough to be
insignificant.}
While the modeling of particular, young SNRs depends critically on the
ejecta and ISM densities \citep[e.g.,][]{DEB2000}, the general
character of the results we present here are insensitive to these
details.  
Specifically, we begin at some time $\tstart < 0.1\, \tch$ with
undisturbed ejecta and ISM separated by a contact discontinuity.
Here, $\tch = \Rch/\Vch$ is the characteristic age with $\Rch = [3
\Mej / (4 \pi \rho_0)]^{1/3}$, $\Vch = \sqrt{2 \EnSN / \Mej}$, and
$\rho_0 = (1 + 4 \fHe) m_p \npz$, where $\EnSN$ is the explosion
energy, $\Mej$ is the ejecta mass, $\npz$ is the ISM proton number
density, $\fHe$ is the helium to proton number ratio, and $m_p$ is the
proton mass.\footnote{Throughout this paper the subscript 0 (2) implies
values upstream (downstream) from the shock.}

The magnetic field, $B$, is ignored in our hydrodynamic model (we
implicitly assume that $B^2/8\pi$ is small compared to the thermal
pressure), but it is an important parameter for the particle
acceleration discussed next.

\subsection{Nonlinear Diffusive Shock Acceleration}
The full details of the nonlinear acceleration model used here are
given in \citet{BE99} and \citet{EBB00}. This is an approximate,
algebraic model of diffusive shock acceleration containing the
essential physics of NL acceleration, but which parameterizes
important properties of the process such as the injection efficiency
and the maximum energy particles achieve.  While more complete models
of nonlinear shock acceleration exist
\citep[e.g.,][]{JE91,BEK96,MD2001}, our algebraic approximation is
easier to include in global models of SNRs. It is computationally fast
making it far less time consuming to do parameter searches and to
compare model results with observations. In Sect.~\ref{sec:Berezhko}
we show by direct comparison that it gives similar results to the more
physically complete model of \citet{BKV2002}.

Briefly, the nonlinear effects in \dsa\ are:
(i) the self-generation of magnetic turbulence by counter-streaming
energetic particles. Back streaming particles produce turbulence in
the magnetic field which leads to stronger scattering of the particles
and hence to more acceleration, quickly leading to saturated
turbulence levels near $\delta B/B \sim 1$ in strong shocks;
(ii) the modification (\iec\ smoothing) of the shock precursor by the
backpressure of energetic particles. The precursor influences the
subshock compression, $\Rsub$, the injection and acceleration
efficiencies, and the shape of the accelerated spectrum.
Since particle diffusion lengths are generally increasing functions of
momentum \citep[\egc][]{BE87,GBSE93}, high momentum particles sample a
broader portion of the flow velocity profile, and hence experience
larger effective total compression ratios, $\Rtot$, than low momentum
particles. Consequently, higher momentum particles have a flatter
power-law index than those at lower momenta and can dominate the
pressure in a NL fashion. The resultant superthermal distribution has
a characteristic concave upward curvature until the spectrum turns
over at the highest energies from losses
\citep[e.g.,][]{EE84,Blasi2002,MDJ2002};
and
(iii) the increase in $\Rtot$ from relativistic particle pressure and
particle escape.
As relativistic particles are produced and contribute
significantly to the total pressure, their softer equation of state
makes the shocked plasma more compressible
($\gamma  \rightarrow 4/3$).  
Even more important, as the highest energy particles escape from
strong shocks they drain away energy flux which must be compensated
for by ramping up the overall compression ratio to conserve the
fluxes.  Just as in radiative shocks, this is equivalent to $\gamma
\rightarrow 1$ and $\Rtot$ can become arbitrarily large
\citep[\egc][]{KE86,BE99,Malkov97}.
As the overall compression increases ($\Rtot > 4$), the subshock
compression ratio, $\Rsub$, which is responsible for heating the gas,
must become less than the test-particle (TP) value ($\Rsub < 4$),
causing the temperature of the shocked gas to drop below TP values.
These changes in shock compression occur simultaneously with changes
in the shape of the accelerated particle spectrum, thus linking X-ray
heating to cosmic-ray production.
For reviews on \dsa\ see \citet{Drury83}; \citet{BE87};
\citet{BereKrym88}; \citet{JE91}; and \citet{MD2001}.

The most important parameters associated with nonlinear shock
acceleration are the Mach numbers (\iec\ the shock speed, $\Vsk$,
pre-shock hydrogen number density, $\npz$, and preshock magnetic field,
$B_0$),
the injection efficiency, $\etainj$ (\iec\ the fraction of
total protons which end up with superthermal energies), and the
maximum proton energy produced, $\Emax$.  As described in
\citet{BE99}, our model includes \alf\ heating in the precursor which
reduces the efficiency compared to adiabatic heating and makes the
magnetic field strength an important parameter.
For given sonic and \alf\ Mach numbers (\iec\ given $M_S=\sqrt{\rho_0
\Vsk^2/(\gamma P_0)}$ and $M_A=\sqrt{4 \pi \rho_0 \Vsk^2}/B_0$), and a
given shock size and age, $\etainj$ sets the overall acceleration
efficiency and determines the importance of NL effects. 
With other parameters fixed, \alf\ wave heating causes the
acceleration efficiency to decrease with increasing $B_0$.
In a complete
model of diffusive shock acceleration, the injection efficiency would
be determined from first principles. However, no current model of
diffusive shock acceleration can do this and injection remains
dependent on approximations of poorly understood wave-particle
interactions.\footnote{While in principle plasma
simulations, where particles move in response to Newton's and
Maxwell's equations (particle-in-cell), can fully describe
wave-particle interactions and injection, in practice, they have not
yet done so because of computational
limits.  The limits have been
insurmountable for two main reasons:
(i) The simulations must be performed fully in 3D because 1- or 2-D
simulations unphysically prevent cross-field diffusion
\citep{JKG93,JJB98}.  In all cases except strictly parallel shocks
(where the upstream magnetic field is parallel to the shock normal),
cross-field diffusion will be an essential part of the injection and
acceleration process;
and (ii) In order for NL effects to become apparent or field
amplification to occur on large scales, the simulations
must be run long enough in a large enough box with enough particles
for a significant population of superthermal particles to be produced.
If, in addition, electrons are to be investigated, the
simulations must use the short electron time-step yet run for many
proton time-scales increasing computation time considerably.
\label{fn:threeD}
}
Here, we investigate the effects of efficient acceleration by varying
$\etainj$.  A principle aim of future work is to constrain $\etainj$
from models using X-ray and broad-band observations of particular
SNRs.  

In order to compare our results directly to those of \citet{BKV2002},
we assume as they do that the magnetic field is turbulent, adopting
the Bohm limit for strong particle scattering, and somewhat
arbitrarily take the field downstream from the shock to be the
compressed upstream magnetic field i.e., $B_2 = \Rtot B_0$, where $B_0
(B_2)$ is the upstream (downstream) magnetic field strength.
We do not expressly consider shock obliquity, i.e., the angle between
the local shock normal and $B_0$, even though this may be an important
factor for understanding emission around the rims of some SNRs
\citep[see][for a discussion of the effects of shock obliquity in a
test-particle description of particle acceleration in
SNRs]{Reynolds98}.  
As a crude approximation, we could model the
asymmetry seen in many SNRs, including SN 1006, by combining
results for different quadrants of the remnant where values of the
magnetic field and injection parameter were varied.

In our examples presented here, we take the unshocked ISM field to be
$B_0 = 20$ \muG\ to match the value determined by \citet{BKV2002} for
SN 1006.
For simplicity, unless explicitly stated we use the same constant value
for the field in the unshocked supernova ejecta even though, in
reality, this field is likely to weaken considerably with time because
of flux conservation [in the discussions associated with
Figs.~\ref{fig:TP_NL_stack} (Sect.~\ref{sec:CR_hydro}) and
\ref{fig:fp_500yr} (Sect.~\ref{sec:spectra}), we show some effects
of a weak ejecta field].

The maximum energy cosmic rays obtain depends, in part, on the
scattering mean free path, $\lambda$, which is assumed to be,
\begin{equation}
\label{eq:mfp}
\lambda = 
\etamfp \, \rg 
\ ,
\end{equation}
where $\etamfp \ga 1$ is taken to be a constant and $\rg= p/(q B)$
is the gyroradius in SI units.  Small values of $\etamfp$ imply strong
scattering and allow higher maximum proton energies in a given system.
The Bohm limit implies $\etamfp \simeq 1$.

%                                                One column figure
   \begin{figure}             % Figure 1
   \centering
   \includegraphics[width=8cm]{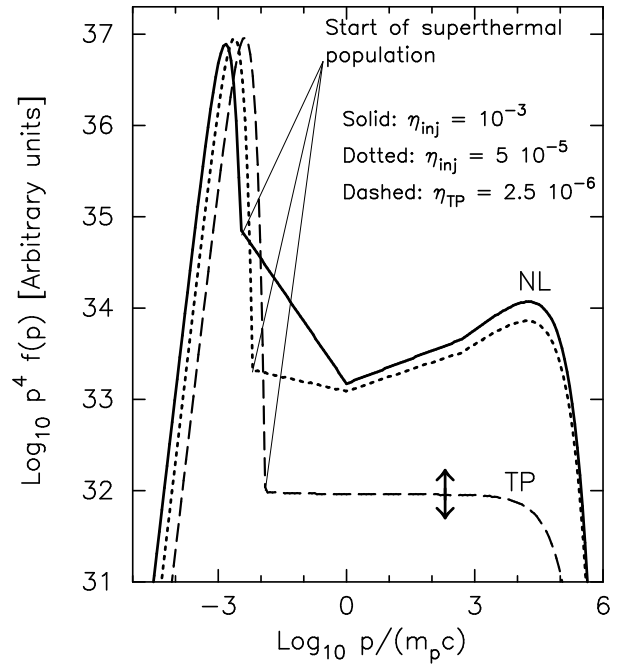}
      \caption{Schematic particle distribution functions, $f(p)$, versus
  momentum, $p$, where $p^4 f(p)$ is plotted to emphasize the spectral
  curvature in the nonlinear results (solid and dotted curves).  The
  kinks in these spectra are manifestations of the piecewise
  approximation used in \citet{BE99}. The lowest momentum kink
  separates the thermal from superthermal population. Above that is
  the three-component power law, and the turnover at the highest
  momentum.  The up and down arrow on the test-particle result (dashed
  curve) indicates that there is no constraint in TP models on the
  density of superthermal particles other than that their energy
  density be insignificant.
}
\label{fig:fp_TP_NL_schmat}
   \end{figure}

The phase-space momentum distributions for protons, $f(p)$, are
calculated as in \citet{EBB00} and consist of a thermal component, a
three-component power law at superthermal energies, and a turnover at
the highest energies given by
\begin{equation}
\label{eq:expo}
\exp{\left [ -\frac{1}{\alpha} 
\left ( \frac{p}{\pmax} \right )^{\alpha}
\right ]}
\ .
\end{equation}
Here, $\alpha$ is a constant and $\pmax = \Emax/c$ is determined by
setting the acceleration time equal to the SNR age, $\tSNR$, or by
setting the diffusion length of the highest energy particles equal to
some fraction, $\fRad$, of the shock radius, $\Rsk$, whichever gives
the lowest $\pmax$ \citep[see][]{BaringEtal99}. 
With these assumptions, $\pmax$ is proportional to the magnetic
field strength.
\citet{Reynolds98} has shown that when 
\synch\ emission is 
summed over space in a spherically symmetric SNR model, the
resultant photon spectrum tends to be broader than that produced by a
single electron distribution falling off as $\exp{(-p/\pmax)}$. This
behavior can be approximated in our model by varying $\alpha$ and
$\alpha \sim 0.5$ has been used to fit X-ray and radio spectra in
particular SNRs \citep[\egc][]{BKP99,ESG2001}. However, 
\citet{BKV2002} obtain a good fit to SN1006 with a sharp turnover in
$f(p)$ and to match their results, we take $\alpha = 4$.

Fig.~\ref{fig:fp_TP_NL_schmat} illustrates the essential differences
in particle spectra between TP shock acceleration (dashed line) and NL
acceleration with different injection efficiencies (solid and dotted
lines).
Compared to the TP power law, i.e.,
\begin{equation}
\label{eq:TPpowerlaw}
f(p) \, d^3 p\propto
p^{-\sigma} \, d^3p
\quad
\hbox{\rm with}
\quad
\sigma = 
\frac{3 \Rtot}{\Rtot -1}
\ ,
\end{equation}
the high-energy portion ($p >m_p c$) of a NL spectrum is flatter, the
low-energy portion of the superthermal spectrum (above the thermal
peak and below $m_p c$) is steeper, and the thermal part is at a lower
temperature.
In the TP approximation, the normalization of the
power-law portion of the spectrum (relative to the thermal peak) is
arbitrary as long as it is low enough to contain an insignificant
fraction of the total energy.
If we arbitrarily set this fraction at 0.01, we can determine, for a
given set of shock parameters, the
injection efficiency, $\etaTP$, below which the thermal gas is
decoupled from the power law, i.e., for any $\etainj \le \etaTP$, the
temperature and density of the thermal gas remains constant.
For the example shown in Fig.~\ref{fig:fp_TP_NL_schmat}, $\etaTP
\simeq 2.5\xx{-6}$.

In contrast, the energetic portions of NL spectra are set by the
conservation of mass, momentum, and energy fluxes and any change in
the relativistic population influences the thermal population.  When
acceleration is efficient, the relation between the shock velocity,
$\Vsk$, and the temperature of the shocked gas, $T_2$, is no longer
approximated by the commonly used expression for a strong, TP shock.
That is,
\begin{equation}
\label{eq:TPtemp}
\frac{k T_2}{\mu \, m_p\Vsk^2}  \ne \frac{3}{16}
\ ,
\end{equation}
where $\mu \, m_p$ is the mean particle mass and $k$ is Boltzmann's
constant. If the equality in equation~(\ref{eq:TPtemp}) is assumed for
a fully ionized plasma with $\nHe/\npz =0.1$ ($\nHe$ is the unshocked
number density of helium) and equal downstream temperatures for all
species,
\begin{equation}
\label{eq:TPtempCon}
T_2 \simeq 1.3\xx{7} \, 
\left ( \frac{\Vsk}{10^3\, {\rm km \, s}^{-1}}\right )^2
{\rm K}
\ .
\end{equation}
For young SNRs, this predicts temperatures just behind the blast wave
well above those capable of explaining observed X-ray line emission
(\egc\ $\Vsk=3000$ \kmps\ $\too$ $T_2\simeq 1.2\xx{8}$ K) and suggests
that the shocked {\it electron} temperature is considerably less than
predicted by equation~(\ref{eq:TPtempCon})
\citep[e.g.,][]{HRD2000,HDHP2002,Long2003}.  

Normally, it is assumed that electrons obtain a much lower temperature
than the ions in the shock layer and take some time to equilibrate
downstream through Coulomb collisions \citep[see][for a discussion of
electron-ion equilibration in modified shocks]{DE2001}. However, the
heating process in the shock layer is certainly dominated by
collisionless, wave-particle interactions which are poorly
understood. The simplest possibility, that all particles upon crossing
the shock gain a speed differential $\Delta u \sim u_0 - u_2$, where
$u_2$ is the downstream flow speed as seen from the shock, predicts
that ions are heated far more than electrons.\footnote{That the
situation is clearly more complicated than this is evident from the
fact that if both species only receive $\Delta u$ upon crossing the
shock, energy is not conserved \citep[see][]{JE87}.}
Nevertheless, the possibility exists that electrons equilibrate faster
than this simple conjecture implies and the fact that efficient
acceleration produces a considerably lower proton temperature
(Fig.~\ref{fig:fp_TP_NL_schmat}) in strong shocks than the TP
prediction of equation~(\ref{eq:TPtempCon}), suggests that it may not
be as obvious that $\Tetwo/\Tptwo \ll 1$ immediately behind the
shock.\footnote{Estimates of $\Tetwo/\Tptwo$ based on optical and UV
line observations have been obtained just behind a few SNR shocks
\citep[e.g.,][]{Raymond_etal95,Laming96,Ghavamian99}. These show a
wide range, i.e., $0.2 < \Tetwo/\Tptwo < 0.8$, with large
uncertainties and are generally restricted to regions that are
partially neutral where the most efficient particle acceleration may
not be occurring \citep[see][for a more complete
discussion]{Drury_issi2001}. }

Furthermore, radio \synch\ observations have long confirmed
that many SNRs accelerate electrons to relativistic energies,
although it is not certain that these electrons are drawn from the
shock heated population in all cases.
What is certain is that if relatively cold upstream electrons enter
the downstream region without being heated and then slowly equilibrate
as they advect away downstream, very few if any will be injected into
the Fermi acceleration process. In the simplest, kinematic models of
injection in Fermi acceleration \citep[e.g.,][]{JE91,BaringEtal99},
the efficient acceleration of thermal electrons is inconsistent with a
lack of electron heating in the shock layer.

\begin{figure}              % Figure 2
   \centering
   \includegraphics[width=8cm]{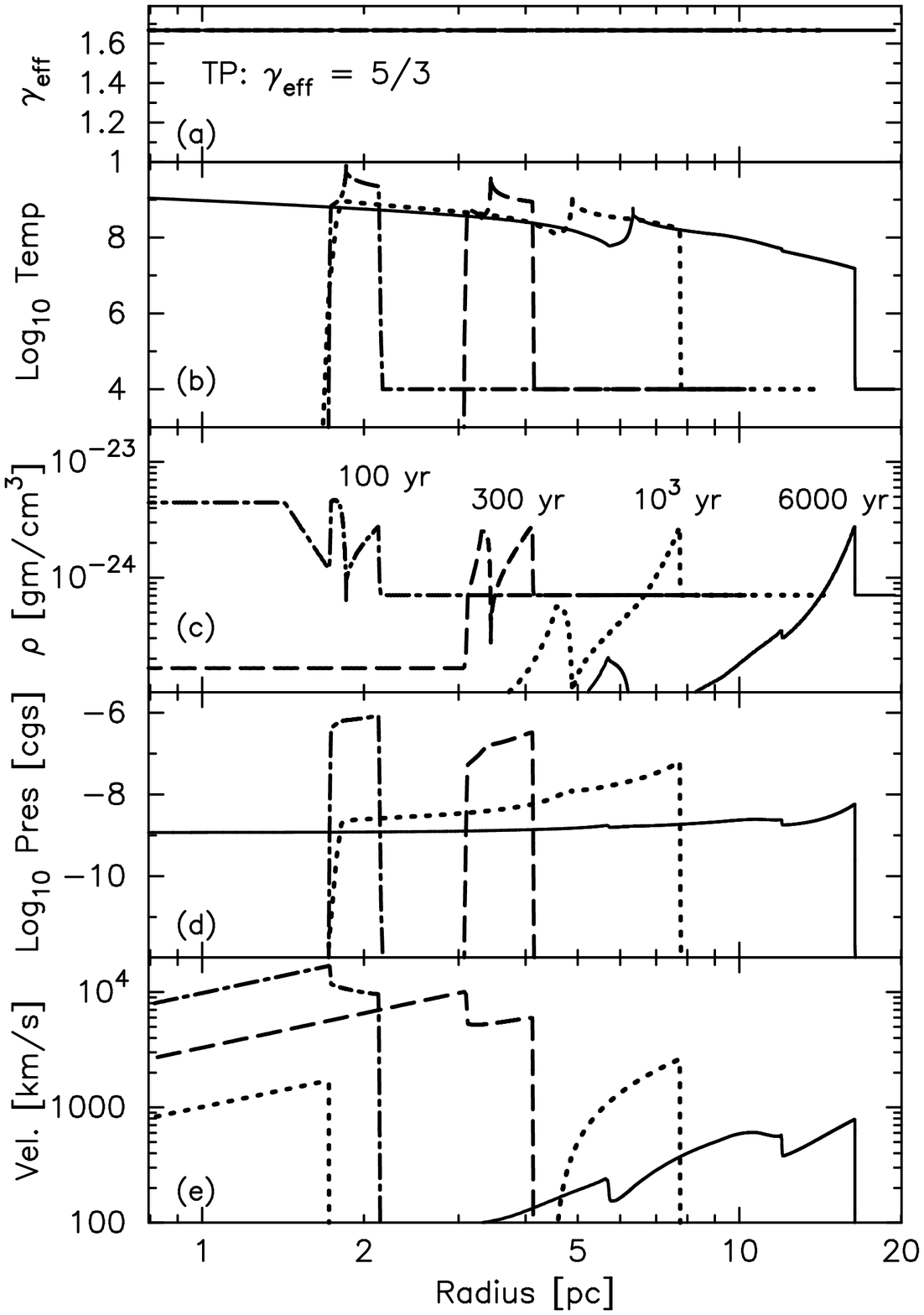} 
\caption{Parameters from the hydro simulation versus radius at
various times during the simulation as labeled. These are
test-particle results with no adjustment of $\gameff$ from particle
acceleration. We have assumed $\EnSN=3\xx{51}$ erg, $\Mej=1.4 \,
\Msun$, and a constant density ISM with $\npz=0.3$ \pcc, and
$\fHe=0.1$. The ejecta density structure is initially a power-law with
index $n=7$ beyond the plateau at small radii (panel $c$).
\label{fig:prof_TP}
}
\end{figure}

\subsection{CR--Hydrodynamic Coupling}
\label{sec:CR_hydro}
When efficient particle acceleration shifts energy from the thermal
gas into relativistic particles, the fraction of total energy density
in pressure is lowered and the evolution of the SNR is modified.  The
modified evolution, in turn, modifies the shock parameters which
determine the acceleration. We model this coupling by including \NL\
shock acceleration in a standard 1-D hydrodynamic simulation as
follows.

\begin{figure}              % Figure 3
   \centering
   \includegraphics[width=8cm]{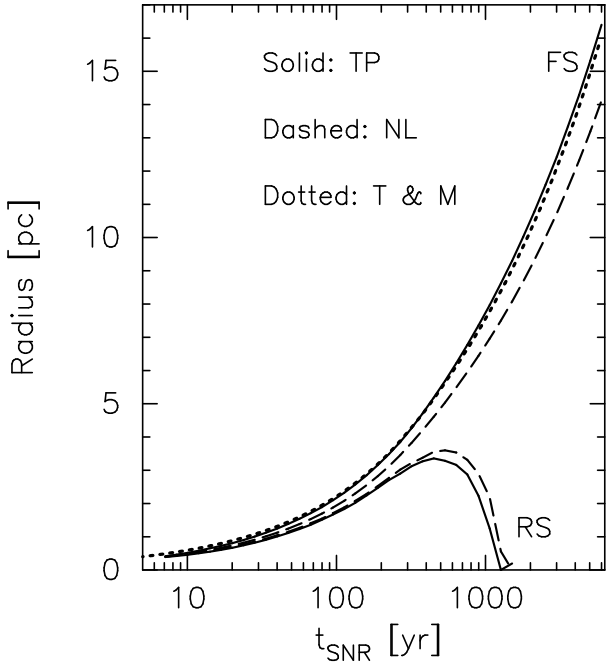} 
\caption{Shock radius versus SNR age for forward (FS) and reverse
shocks (RS) obtained with our hydrodynamic simulation.  The dashed
curves show NL results with efficient particle acceleration
($\etainj=10^{-3}$), while the solid curves are test-particle (TP)
shocks with no particle acceleration.  The dotted curve shows the TP
forward shock radius obtained with the approximate analytic expression
given in \citet{Truelove99}.
\label{fig:rad_vs_age}
}
\end{figure}

\newlistroman

Briefly:
\listromanDE the hydro simulation is initiated at some time, $\tStart
< 0.1 \, \tch$, after the supernova explosion with a plateau-power-law
ejecta distribution and a constant interstellar medium density,
$\rhoISM$;
\listromanDE At each time step, the hydro evolves and the radii and
Mach numbers of the forward and reverse shocks are determined;
\listromanDE With this shock information, the CR acceleration is
calculated using the approximate model of \citet{EBB00} with the
injection parameter, $\etainj$, kept constant during the simulation
and having the same value at the forward and reverse
shocks;\footnote{In an actual SNR, of course, the injection efficiency
might vary with time, vary over the shock surface, or be different at
the forward and reverse shocks \citep[as in our model of Kepler's
SNR;][]{DEB2000}. For the general results presented here, keeping
$\etainj$ constant avoids unnecessary complexity. It is important to
note, however, that the {\it acceleration} efficiency, i.e., the
fraction of ram kinetic energy that is transferred to superthermal
particles, depends on the shock parameters (Mach numbers, shock age,
etc.) as well as on $\etainj$ and does vary with age and between the
forward and reverse shocks even if $\etainj$ is constant.}
\listromanDE 
The acceleration calculation provides the overall compression ratio,
$\Rtot$, and the effective ratio of specific heats is determined by
\begin{equation}
\label{eq:gammaeff}
\gameff = \frac{M_S^2 (\Rtot + 1) - 2 \Rtot}{M_S^2 (\Rtot - 1)}
\ ,
\end{equation}
where we note that $\gameff < (P/\EnDen) + 1$ since $\gameff$ includes
the effects from ``escaping'' particles.\footnote{Performing the
acceleration calculation at each time step is computationally
expensive. If the full spectrum is not needed and only efficient
acceleration is considered, a faster approximation can use $\Rtot =
1.3 M_A^{3/8}$ if $M_A < M_s^2$ or $\Rtot = 1.5 M_s^{3/4}$ otherwise
\citep[see][]{BE99}.}
Here, $P$ ($\EnDen$) is the pressure (energy density) of the shocked
gas;
\listromanDE The coupling between the CR acceleration and the
hydrodynamics is accomplished by replacing the ratio of specific heats
used in the hydrodynamic equations with $\gameff$ in the shells
spanning the shock transition. We use a Lagrangian
mode in the hydro simulation so the simulation grid moves with the
mass. Once a mass element has been shocked and a $\gameff$ has been
assigned to it using equation~(\ref{eq:gammaeff}), that mass element
retains that $\gameff$ for the rest of the simulation or until it
reencounters a shock. Even though energy is leaving the system via the
highest energy cosmic rays, we do not explicitly remove energy from
the hydro. Lowering $\gameff$ causes the gas to be more compressible
and mimics the escaping energy in one step;
\listromanDE As the SNR evolves, the proton distribution functions in
momentum, $f(p)$, associated with each shocked shell undergo adiabatic
losses; and finally,
\listromanDE At the end of the simulation, the $f(p)$'s from all of
the shells are summed to determine the total contribution to the
Galactic CR source population.

To start, we show in Fig.~\ref{fig:prof_TP} a TP case where $\gameff
= 5/3$ always (top panel).  For this example and all others presented
in this paper, we have taken $\EnSN=3\xx{51}$ erg, $\Mej=1.4 \,
\Msun$, $B_0=20$ \muG, and a constant density ISM with $\npz =0.3$
\pcc, and $\fHe=0.1$. These values give a characteristic age, $\tch =
\Rch/\Vch \simeq 210$ yr.  The initial ejecta density distribution is
a power law ($\rho \propto r^{-7}$) with a flat plateau at small
radii.
The simulation spans $6000$ yr and various quantities are shown, as a
function of radius, at different times during the SNR evolution. The
temperature shown in panel $(b)$ is the average temperature defined as
\begin{equation}
<\! T_2 \! > = \frac{P}{\rho} \frac{\mu \, m_p}{k}
\times \facTh
\ ,
\end{equation}
where $P$ is the pressure, $\rho$ is the density, $\mu = (1 + 4
\fHe)/(2 + 3 \fHe)$, $\fHe=0.1$ is the ratio of helium atoms to
protons, and $\facTh$ is the fraction of pressure in thermal
particles. In this plot and all others, we assume the plasma to be
fully ionized.
When acceleration occurs, superthermal particles contribute to the
total hydrodynamic pressure, $P$. We define the temperature from that
fraction of total pressure in the ``thermal'' particles,
$\facTh$. This fraction is determined from $f(p)$ (see
Fig.~\ref{fig:fp_TP_NL_schmat} where the division between thermal
and superthermal particles is indicated).

\begin{figure*}              % Figure 4
   \centering
   \includegraphics[width=12cm]{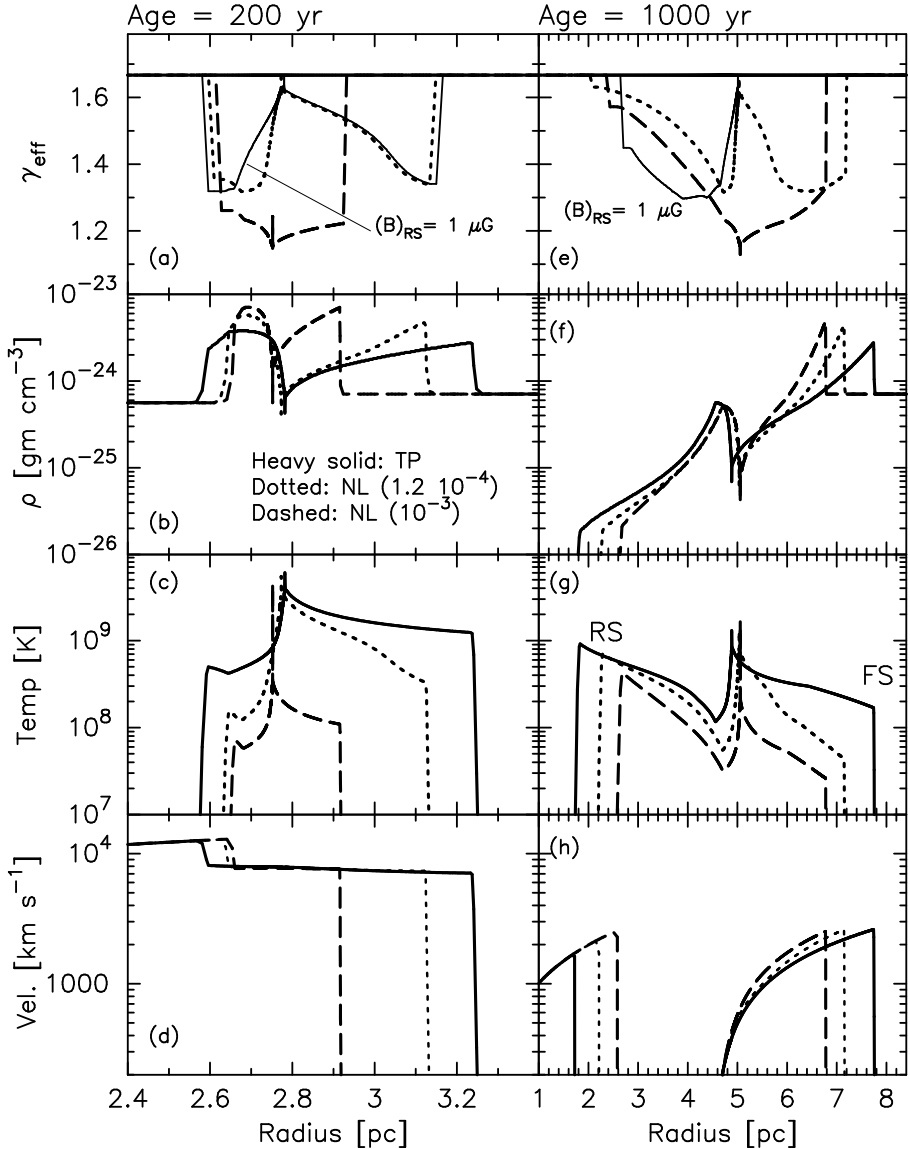} 
\caption{Comparison of the parameters for a test-particle
simulation (solid curves) with those undergoing particle acceleration.
The results are shown at $\tSNR=200$ and 1000 yr and the supernova
parameters are the same as those in Fig.~\ref{fig:prof_TP}.  The
light-weight solid curves in panels (a) and (e) show the effect on
$\gameff$ of reducing the unshocked ejecta magnetic field from
$(B_0)_\mathrm{RS}=20$ \muG\ to 1 \muG.  Apart from
$(B_0)_\mathrm{RS}$, all other parameters are the same as in the
$\etainj=1.2\xx{-4}$ examples shown with dotted curves. In panel (e),
the shocked ISM portion of the $(B_0)_\mathrm{RS} = 1$ \muG\ curve is
essentially identical to the dotted curve.
\label{fig:TP_NL_stack}
}
\end{figure*}

For the particular parameters assumed for the model shown in
Fig.~\ref{fig:prof_TP}, the SNR is still in the self-similar phase
after 100 yr as the reverse shock moves through the power-law portion
of the ejecta. By 300 years, the reverse shock is moving through the
plateau region of the ejecta and the forward shock has reached $\radFS
\simeq 4$ pc. The density of the unshocked ejecta is now well below
the ambient ISM density $\rhoISM=1.4 \, m_p \npz$ at this stage. By
1000 years, $\radFS \sim 8$ pc, the RS is now moving back toward the
origin, and the energy of the explosion is shifting from being in the
kinetic energy of the ejecta to thermal pressure.  By $6000$ years,
the RS has collapsed back through the ejecta to the origin and the SNR
is well into the Sedov phase with a FS speed of about $(4/3) 750$
\kmps.\footnote{Note that the speed shown in the bottom panel of
Fig.~\ref{fig:prof_TP} and panel $(d)$ of
Fig.~\ref{fig:TP_NL_stack} is the flow speed in the frame of the
explosion. The shock speed is given by $\Vsk = (\Rtot \uDS -
\uUP)/(\Rtot - 1)$, where $\uUP$ ($\uDS$) is the flow speed just
upstream (downstream) from the shock and $\Rtot$ is the shock
compression ratio. For the ISM conditions we assume here, $\uUP$
always equals zero for the forward shock. At $\tSNR=6000$ yr in
Fig.~\ref{fig:prof_TP} at the FS, $\uDS\simeq 750$ \kmps, $\Rtot
\simeq 4$, and $u_0 \simeq 1000$ \kmps.}

In Fig.~\ref{fig:rad_vs_age} we illustrate how particle acceleration
influences the SNR shock evolution by comparing TP results ($\etainj
\la 10^{-6}$) against a simulation with $\etainj=10^{-3}$. The
figure shows the forward and reverse shock radii versus remnant age,
$\tSNR$, for the same parameters as used in
Fig.~\ref{fig:prof_TP}.\footnote{We do not follow particle
acceleration at the reverse shock after it first collapses to zero
radius.} 
The TP results, i.e., those with $\gameff=5/3$ everywhere,
are shown with solid curves and the NL results are shown with dashed
curves. The dotted curve is the TP analytic approximation given by
\citet{Truelove99} for the FS and this matches the hydro results well
over the entire time period.
When NL effects are important, the region between the forward and
reverse shocks becomes narrower and denser, a result fully consistent
with the analytic results of \citet{DEB2000} and the multi-dimensional
hydro simulations of \citet{BE2001}.

In Fig.~\ref{fig:TP_NL_stack} we compare TP profiles to NL ones at
$\tSNR = 200$ and 1000 yr. In all panels, the heavy-weight
solid curves are the TP
results ($\etainj \la 10^{-6}$), the dotted curves have
$\etainj=1.2\xx{-4}$, and the dashed curves have $\etainj=10^{-3}$.
Panels ($a$) and ($e$) show the reduction in $\gameff$ that occurs as
shock acceleration shifts energy into relativistic particles. This
produces compression ratios greater than 4 and results in denser,
narrower regions between the forward and reverse shocks as shown in
the density profiles ($b$) and ($f$). 
The separation between the forward and reverse shocks varies
considerably with $\tSNR$ and $\etainj$. At $\tSNR=200$ yr, this
separation is $\sim 0.25$ pc for $\etainj=10^{-3}$ and $\sim 0.7$ pc
for the TP case, while at $\tSNR=1000$ yr it is $\sim 4.2$ pc for
$\etainj=10^{-3}$ and $\sim 6$ pc for the TP case.  The shift in
energy out of the thermal gas also results in a significant decrease
in the temperature of the shocked gas in the NL case, as shown in
panels ($c$) and ($g$).

The flow speed is shown in the bottom panel and the forward and
reverse shocks are easily identified in all panels. 
At $\tSNR=1000$ yr, the flow speeds just behind the forward shocks of
the three examples are about equal, i.e., $\uDS({\rm TP}) \simeq 2600$
\kmps\ and $\uDS(\etainj=10^{-3}) \simeq 2500$ \kmps. However, since
the shock speed depends on $\Rtot$, the TP shock is moving faster,
i.e., $u_0({\rm TP}) \simeq (4/3) 2600 \simeq 3500$ \kmps, while
$u_0(\etainj=10^{-3}) \simeq (7/6) 2500 \simeq 2900$ \kmps.

The intermediate NL case with $\etainj=1.2\xx{-4}$ (dotted curves)
produces temperatures and compression ratios between the TP and
$\etainj=10^{-3}$ cases, as expected. However, as seen in the
variations of $\gameff$ (panels $a$ and $e$ of
Fig.~\ref{fig:TP_NL_stack}), the $\etainj=1.2\xx{-4}$ case acts
quite differently at the beginning of the simulation.  As described in
\citet{BE99}, {\it unmodified} shocks with $\Rtot \simeq 4$ can occur
for high Mach numbers if $\etainj$ is small enough. In this case, the
pressure in relativistic particles, $\Prel$, is small compared to
$\rho_0 \Vsk^2$ and these particles do not slow the incoming gas
enough to produce the \NL\ shock modification.  For a given $\etainj$,
as the SNR ages, $\Vsk$ decreases, $\Prel/(\rho_0 \Vsk^2)$ increases,
and the shock acceleration becomes more efficient and more \NL. Thus,
for relatively inefficient injection, $\gameff$ drops and $\Rtot$
increases in the early stages of evolution, exactly the opposite
behavior as with efficient injection.\footnote{As discussed in
\citet{BE99}, it is by no means certain that actual SNR shocks will
have injection rates low enough for high Mach number, unmodified
shocks to occur. Furthermore, there may be large differences in
injection rates for parallel and oblique regions of the same SNR blast
wave, making it possible that some regions are highly modified and
others unmodified.}

After the initial transition from a strong, unmodified shock to a
strong, modified one, the forward shock will enter the long stage
where it slowly weakens, the acceleration becomes less efficient, and
$\gameff$ increases toward 5/3.
However, as shown by the dotted curve in panel $(e)$ in
Fig.~\ref{fig:TP_NL_stack}, the reverse shock weakens much more
quickly and at $\tSNR=1000$ yr has $\gameff \simeq 5/3$. Since we
assume that $B_0$ is a constant everywhere, the RS weakens as the
ejecta density drops and the magnetic pressure becomes dominant over
the gas pressure.
In an actual SNR, magnetic flux conservation in the expanding,
unshocked ejecta would cause $B_0$ to weaken as the ejecta density
drops and the RS would remain stronger than with a constant $B_0$.
For a given unshocked density, a weak $B_0$ results in efficient shock
acceleration because the \alf\ Mach number increases with decreasing
$B_0$, and because the amount of energy transferred out of accelerated
particles into waves and then into heating of the precursor gets less.
Of course, if the magnetic field gets too weak, the gyroradii of
accelerated particles will become comparable to the shock radius, or
the acceleration time will become comparable to $\tSNR$, before high
particle energies or high efficiencies are obtained.  To illustrate
the effects of a low ejecta magnetic field, we show in panels (a) and
(e) of Fig.~\ref{fig:TP_NL_stack} $\gameff$ produced by our
$\etainj=1.2\xx{-4}$ example with the unshocked ejecta field set at 1
\muG\ (light-weight solid curves).  As seen clearly in panel (e),
$\gameff$ remains lower at the end of the simulation than in the
$(B_0)_\mathrm{RS} = 20$ \muG\ case, indicating more efficient
acceleration.

\begin{figure}              % Figure 5
   \centering
   \includegraphics[width=8cm]{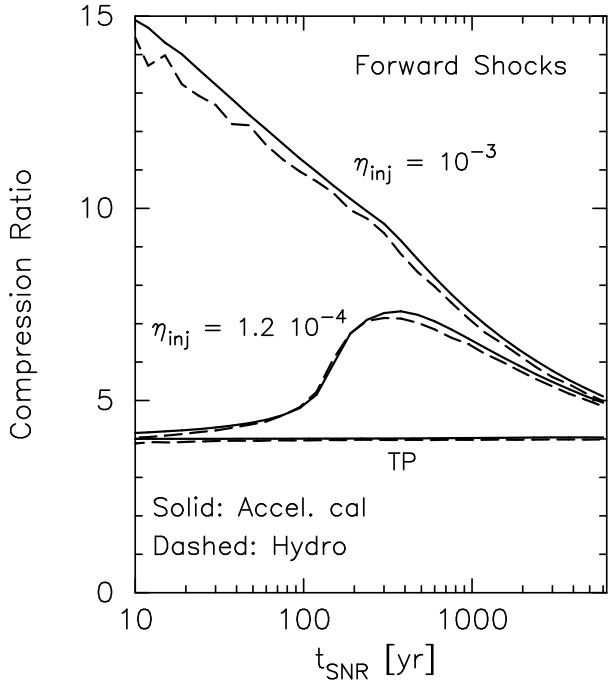} 
\caption{Forward shock compression ratios vs. supernova age for
various injection rates, $\etainj$. In all cases, the solid curves are
obtained from the NL particle acceleration calculation and the dashed
curves are obtained from the density ratio, $\rhoDW/\rhoUP$, just
downstream and upstream from the shock.
\label{fig:Rtot}
}
\end{figure}

In Fig.~\ref{fig:Rtot} we plot the compression ratio, $\Rtot$,
versus $\tSNR$ for our TP and two NL examples. There are
two sets of curves because there are two ways of determining
$\Rtot$. The solid curves are values obtained from the particle
acceleration calculation which are then used in
equation~(\ref{eq:gammaeff}) to calculate $\gameff$. The dashed curves
are determined directly from the hydro density ratio taken just
downstream and upstream from the shock, $\rhoDW/\rhoUP$. In principle,
these should be identical but in practice differences occur,
indicating the inherent errors in our $\gameff$ technique. The largest
average difference shown here with $\etainj=10^{-3}$ is $< 10\%$.
The $\etainj=1.2\xx{-4}$ case clearly shows the rapid transition from the
strong, unmodified shock with $\Rtot \sim 4$ at early times to the
modified shock with $\Rtot > 4$ at intermediate times. The striking
difference between the $\etainj=10^{-3}$ and $\etainj=1.2\xx{-4}$ cases
will leave an imprint on the post-shock gas that may be an
important diagnostic for determining if efficient Fermi acceleration
takes place in young SNRs.

\begin{figure}              % Figure 6
   \centering
   \includegraphics[width=8cm]{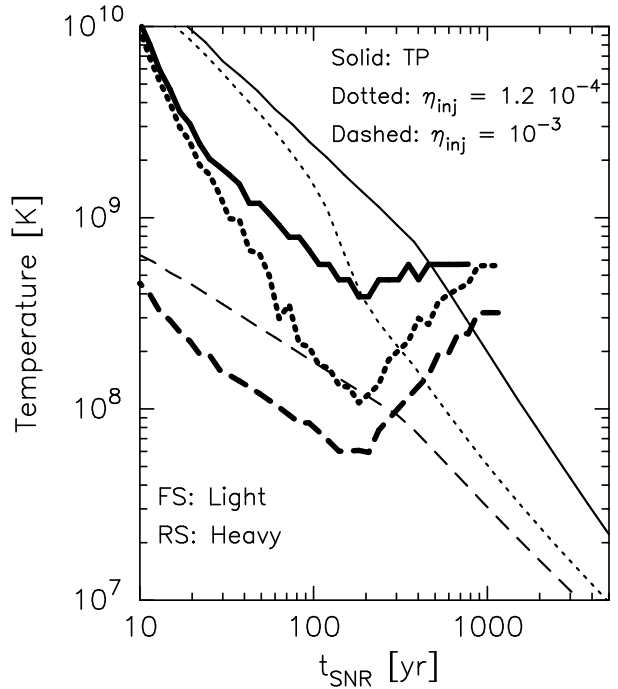} 
\caption{Temperatures immediately behind the forward and reverse
shocks as a function of $\tSNR$. The solid curves are the TP case, the
dotted curves are for $\etainj=1.2\xx{-4}$, and the dashed curves are
for $\etainj=10^{-3}$. The reverse shock temperatures are plotted with
heavy-weight curves.
\label{fig:temp}
}
\end{figure}

In Fig.~\ref{fig:temp} we show the postshock proton temperatures for
the same three examples. Comparing the heavy-weight solid and dashed
curves shows that the RS temperature in the $\etainj=10^{-3}$
efficient shock is less than $\sim 1/10$ that of the test-particle
shock over most of the time period shown. This is a difference large
enough to have profound consequences for the interpretation of X-ray
thermal emission in young SNRs. Furthermore, the temperature in the
intermediate case ($\etainj=1.2\xx{-4}$) shows a stronger time
variation than either the TP or $\etainj=10^{-3}$ case. 
This is reflected as a somewhat steeper spatial gradient in
temperature as seen in panel $(g)$ of Fig.~\ref{fig:TP_NL_stack},
again offering a possible diagnostic to test for efficient particle
acceleration.

\subsubsection{Comparison with Analytic Model of Decourchelle et al.}
In our previous work \citep[][]{DEB2000} we coupled NL acceleration
with an analytic, self-similar description of the SNR hydrodynamics
\citep[i.e.,][]{Chev83} and a non-equilibrium ionization calculation
of thermal X-ray emission. That model used the same algebraic
acceleration calculation we use here with the same injection
parameter, $\etainj$.
The assumptions of the two models differ mainly in the way the
hydrodynamics are treated.

\begin{figure}              % Figure 7
   \centering
   \includegraphics[width=8cm]{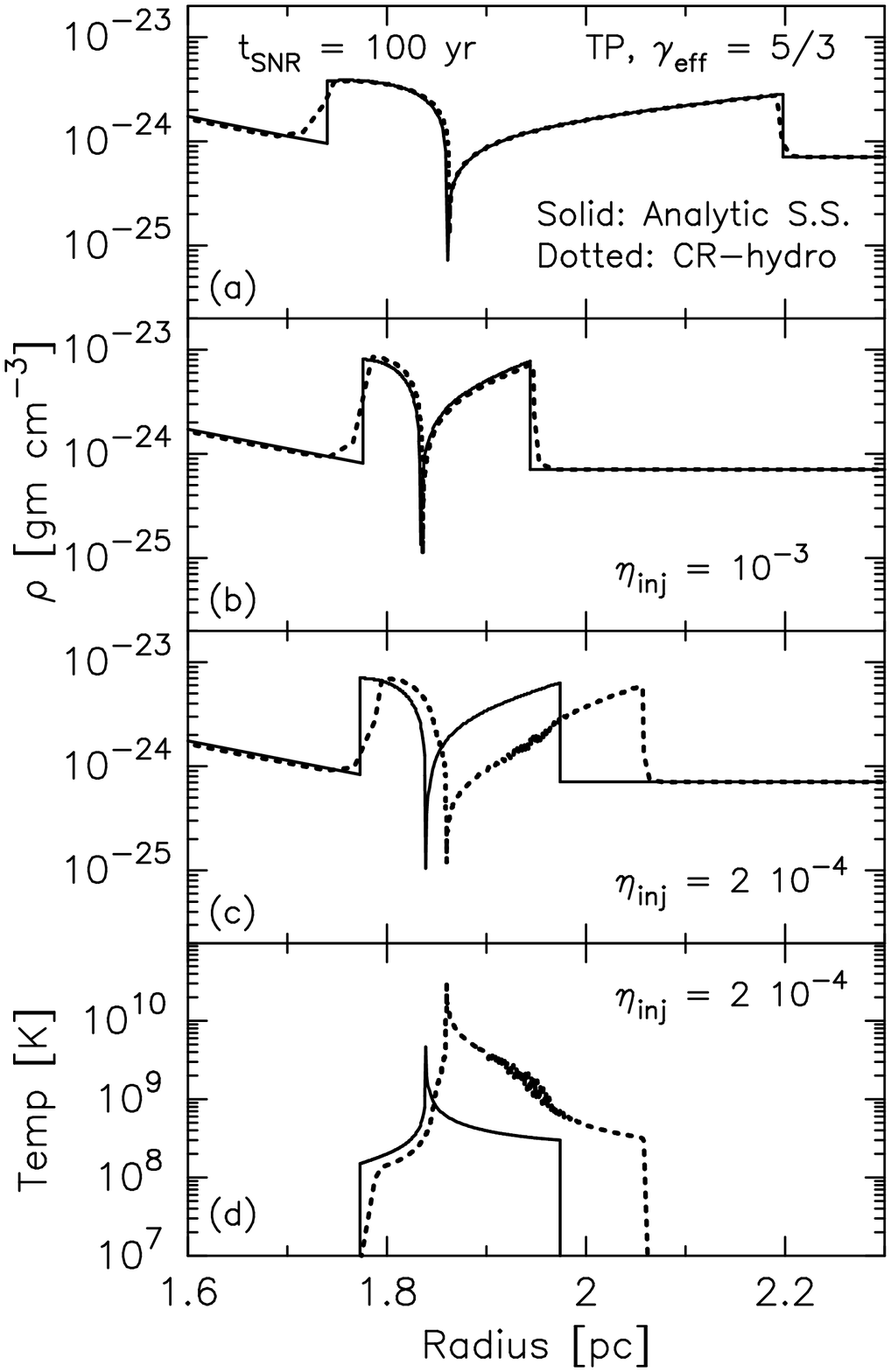} 
\caption{Density (top three panels) versus radius for two NL models
and a TP model, as indicated, at $\tSNR=100$ yr.  The bottom panel
shows temperature versus radius for $\etainj=2\xx{-4}$.  In all
panels, the solid curves are calculated using the analytic,
self-similar model of \citet{DEB2000} and the dotted curves are from
the CR-Hydro model presented here. In the CR-Hydro model, some
numerical noise is evident in the shocked ISM profiles for
$\etainj=2\xx{-4}$.
\label{fig:SS_comp}
}
\end{figure}

While the hydro simulation has a one-fluid approach with an effective
ratio of specific heats derived from the compression ratio
(equation~\ref{eq:gammaeff}), the analytical, self-similar solutions
are based on a two-fluid approach: a nonrelativistic one with
$\gamma=5/3$ and a relativistic one with $\gamma=4/3$.
The NL particle acceleration calculation provides the compression
ratio and the fraction of total pressure in relativistic gas at both
shocks which are used to determine the boundary conditions of the
self-similar solution. Imposing the compression ratios, which can be
greater than 7, mimics the effects of particle loss.
The limitations of the self-similar calculation are that the RS
must be propagating in the power-law portion of the ejecta density
profile, and that it assumes that the forward and reverse shock
compression ratios, as well as the fractions of total pressure in
relativistic gas, remain constant throughout the evolution.
A limitation of our hydrodynamic simulation is that it assigns 
at the shocks to an element of gas an effective gamma, which is kept 
constant during its post-shock evolution.

In the top three panels of Fig.~\ref{fig:SS_comp} we compare the
density structures for our two models at $\tSNR=100$ yr when the RS is
still propagating in the power-law portion of the ejecta density
profile. We use the same supernova and ISM parameters as in our
previous runs.
For the TP (panel $a$) and $\etainj=10^{-3}$ (panel $b$) cases, the
correspondence between the two models is extremely close with small
differences resulting mainly from the finite grid used in the hydro
simulation. 
The reason for this comes from the fact that self-similarity is
satisfied in the TP case where $\Rtot \simeq 4$ always. It is also
approximately satisfied in the $\etainj = 10^{-3}$ case since
$\Rtot$, and therefore $\gameff$, remains fairly constant during the
100 yr evolution, as indicated by the dashed curve in panel $(a)$ of
Fig.~\ref{fig:TP_NL_stack}.

While it was not obvious that the $\gameff$ procedure used in the
hydro model would be a good approximation, the excellent agreement we
see with the analytic model, which only assumes self-similarity,
justifies this approach, at least at these early times.

Much larger differences in the two models are seen in panels $(c)$ and
$(d)$ where $\etainj=2\xx{-4}$.
In this case, $\Rtot$ and therefore $\gameff$ vary
strongly during 100 yr (similar to the dotted curve in panel $a$ of
Fig.~\ref{fig:TP_NL_stack}), and self-similar conditions no longer
apply.
Since the analytic model takes $\Rtot$ at the forward and reverse
shocks at the final time and applies these values for the entire
evolution, it is effectively using values, at least for the FS in the
$\etainj=2\xx{-4}$ case, which are considerably greater than the
average values used in the hydro calculation.
Using a larger $\Rtot$ means there is less pressure for a given energy
density pushing the shock in the self-similar model than in the hydro
model.
Therefore, the FS in the hydro calculation travels
faster and extends further than in the analytic calculation. The
variation in $\gameff$ is less in the shocked ejecta (again similar
to the dotted curve in panel $a$ of Fig.~\ref{fig:TP_NL_stack}) so
the difference in the two models is less there as well.

We consider the close correspondence between these two models in the
TP and $\etainj=10^{-3}$ cases to be a clear indication that both
adequately model the SNR hydrodynamics during self-similar conditions.
The validation of the $\gameff$ approach by our two-fluid
description of the post-shock flow in the framework of self-similar,
cosmic-ray modified Chevalier solutions means the more general hydro
simulation can be used with some confidence when self-similar
conditions do not apply. This will also allow us to use the analytic
model to test the hydro model when X-ray thermal and broad-band photon
emission is included.

\subsection{Energetic Proton Spectra}
\label{sec:spectra}
\begin{figure}              % Figure 8
   \centering
   \includegraphics[width=8cm]{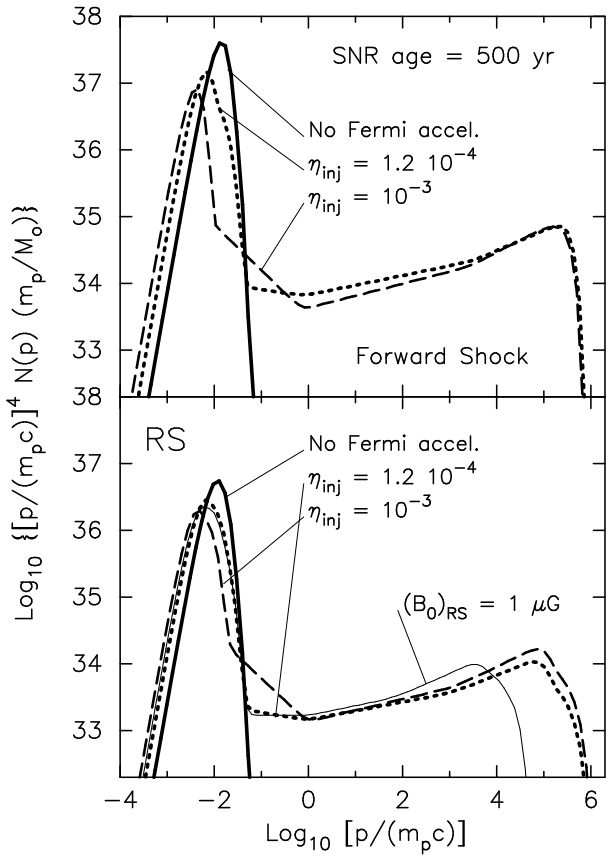} 
\caption{Proton number distributions, $N(p)$ (multiplied by
$[p/(m_pc)]^4 \, m_p/ \Msun$), for hydro simulations with no
acceleration (heavy solid curves), with $\etainj=1.2\xx{-4}$ (dotted
and thin solid curves),
and with $\etainj=10^{-3}$ (dashed curves) summed over 500 years. The
spectra in the top panel are from the shocked ISM (forward shock),
while those in the bottom panel are from the shocked ejecta (reverse
shock). The $N(p)$'s are absolutely normalized to the total number of
protons overtaken by the shock during their lifetime. All spectra have
been adjusted for adiabatic expansion.  
The thin solid curve in the bottom panel shows the RS spectrum
for the case where the unshocked ejecta magnetic field is set at
$(B_0)_{\mathrm{RS}}=1$ \muG.
\label{fig:fp_500yr}
}
\end{figure}

The proton distribution functions that are calculated at the forward
and reverse shocks during the SNR evolution can be summed to determine
the total contribution to the cosmic-ray spectrum.
The amount of material swept up by each shock is used to weight each
spectrum and convert $f(p)$ to $\Ndt$ where $4 \pi \int_0^{\infty} p^2
\Ndt dp$ is the total number of particles overtaken by the shock in
the time interval, $\Delta t$. The $\Ndt$'s are then added together,
with each one adjusted for adiabatic losses during the time from when
the spectrum was calculated until the end of the simulation
\citep[e.g.,][]{Reynolds98}, to produce $N(p)$.\footnote{We do not
include radiation (i.e., \synch) losses here since they are always
negligible for ions. For electrons, of course, these losses must be
considered.}
Typical proton spectra calculated with $B_0=20$ \muG\ at $\tSNR = 500$
yr ($\tSNR/\tch \simeq 2.4$ for $\EnSN=3\xx{51}$ erg, $\Mej=1.4\,
\Msun$, $\npz =0.3$ \pcc, and $\fHe=0.1$) are shown in
Fig.~\ref{fig:fp_500yr}. 
The NL examples produce cosmic rays at the expense of heating and show
lower temperatures, as indicated by the positions of the thermal
peaks. The dashed curve is the result with the most efficient
injection ($\etainj=10^{-3}$) and this spectrum is the flattest at
relativistic energies (before the turnover at $\pmax$). The spectrum
with $\etainj=1.2\xx{-4}$ is the most distorted from a Maxwellian at
thermal energies due to the summing of spectra which evolve from
unmodified ($\Rtot \simeq 4$) to strongly modified ($\Rtot > 4$), as
discussed above.

To illustrate the effects of a weakening unshocked ejecta
magnetic field, we show in the bottom panel of Fig.~\ref{fig:fp_500yr}
(thin solid curve) $N(p)$ for the case where the unshocked ejecta
field is set at $(B_0)_{\mathrm{RS}}=1$ \muG, a factor of 20 lower
than in the other cases. The unshocked ISM field is kept at 20 \muG\
and $\etainj=1.2\xx{-4}$ for both the forward and reverse shocks.  The
flatness of the superthermal part of the spectrum compared to the
dotted curve indicates that the RS is stronger (i.e., $\Rtot$ is
larger) than when $(B_0)_\mathrm{RS} = 20$ \muG\ (see
Fig.~\ref{fig:TP_NL_stack}). However,  $\pmax$ is well below that in the
$(B_0)_{\mathrm{RS}}=20$ \muG\ cases because the weak field
produces large gyroradii and high momentum particles cannot be
accelerated.  The FS spectrum is only slightly changed by changing
$(B_0)_\mathrm{RS}$ and is not shown.

\begin{figure}              % Figure 9
   \centering
   \includegraphics[width=8cm]{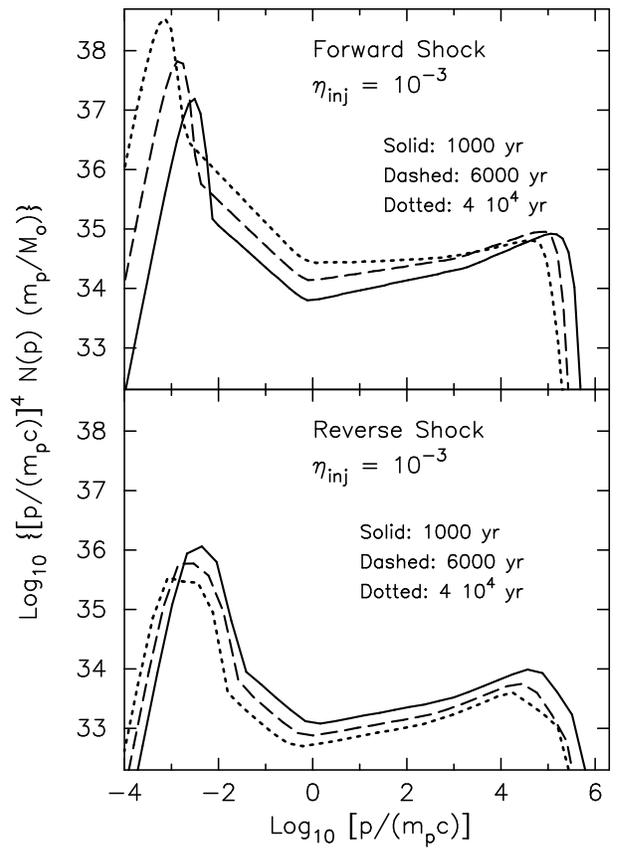} 
\caption{Particle number distributions, $N(p)$ (multiplied by
$[p/(m_pc)]^4 \, m_p/ \Msun$) from the shocked ISM (FS) and shocked
ejecta (RS) at various times during the SNR evolution. Efficient
particle injection is assumed ($\etainj=10^{-3}$) throughout the SNR
evolution, but the fraction of energy in relativistic particles varies
as the SNR ages. The acceleration efficiency decreases as the SNR ages
and the shocks weaken. The RS contributes little to the total CR
population at $\tSNR=4\xx{4}$ yr.  All spectra have been adjusted for
adiabatic expansion.
\label{fig:cr_spec}
}
\end{figure}

In Fig.~\ref{fig:cr_spec} we show spectra at 1000, 6000, and
$4\xx{4}$ yr for the same ISM and SN parameters with
$\etainj=10^{-3}$.  As the FS expands and weakens, the acceleration
efficiency drops and the relativistic portion of the spectrum
steepens. As expected, particles accelerated at the RS decrease in
importance in later stages of the SNR. As we show in
Fig.~\ref{fig:sn1006_comp} below, even moderately efficient
injection (i.e., $\etainj \ga 10^{-4}$) results in approximately
$50\%$ of the explosion energy being put into relativistic particles
over the lifetime of the SNR. Another important property shown in the
top panel of Fig.~\ref{fig:cr_spec} is that, even with
$\etainj=10^{-3}$ which produces strongly curved spectra at early
times, the total particle distribution function after $4\xx{4}$ yr is
approximately $N(p) \propto p^{-2}$ at relativistic energies. Thus the
total cosmic-ray production is fairly consistent with the observed CR
flux after energy dependent escape from the galaxy is included.

\subsection{Comparison with Kinetic Models}
\label{sec:Berezhko}
One of the most complete models of nonlinear CR production in SNRs is
that of Berezhko and co-workers \citep[e.g.,][]{BEK96,BerezV97} where
the time-dependent CR transport equations are solved together with the
gas-dynamic equations in spherical symmetry. This model gives the
radial distribution of gas and accelerated CR spectrum at any phase of
the SNR evolution and has been used with good success to model a
number of young SNRs including SN1006 \citep[][]{BKV2002}, Cassiopeia
A \citep[][]{BPV2003}, and Tycho \citep[][]{VBKR2002}.  In all of the
above cases, the authors conclude that satisfactory fits to the radio
and X-ray \synch\ emission, along with $\gamma$-rays where observed,
are best obtained when the forward shocks are {\it strongly modified}
by the efficient acceleration of protons in a high magnetic field.
While similar results have been presented before
\citep[e.g.,][]{RE92,EBB00,ESG2001}, Berezhko's model probably
contains the most complete physical description of the NL acceleration
process at the FS, and the detailed modeling of particular SNRs
clearly lends credibility to the suggestion that shocks in young SNRs
can be strongly nonlinear with compression ratios greater than 4.

\begin{figure}              % Figure 10
   \centering
   \includegraphics[width=8cm]{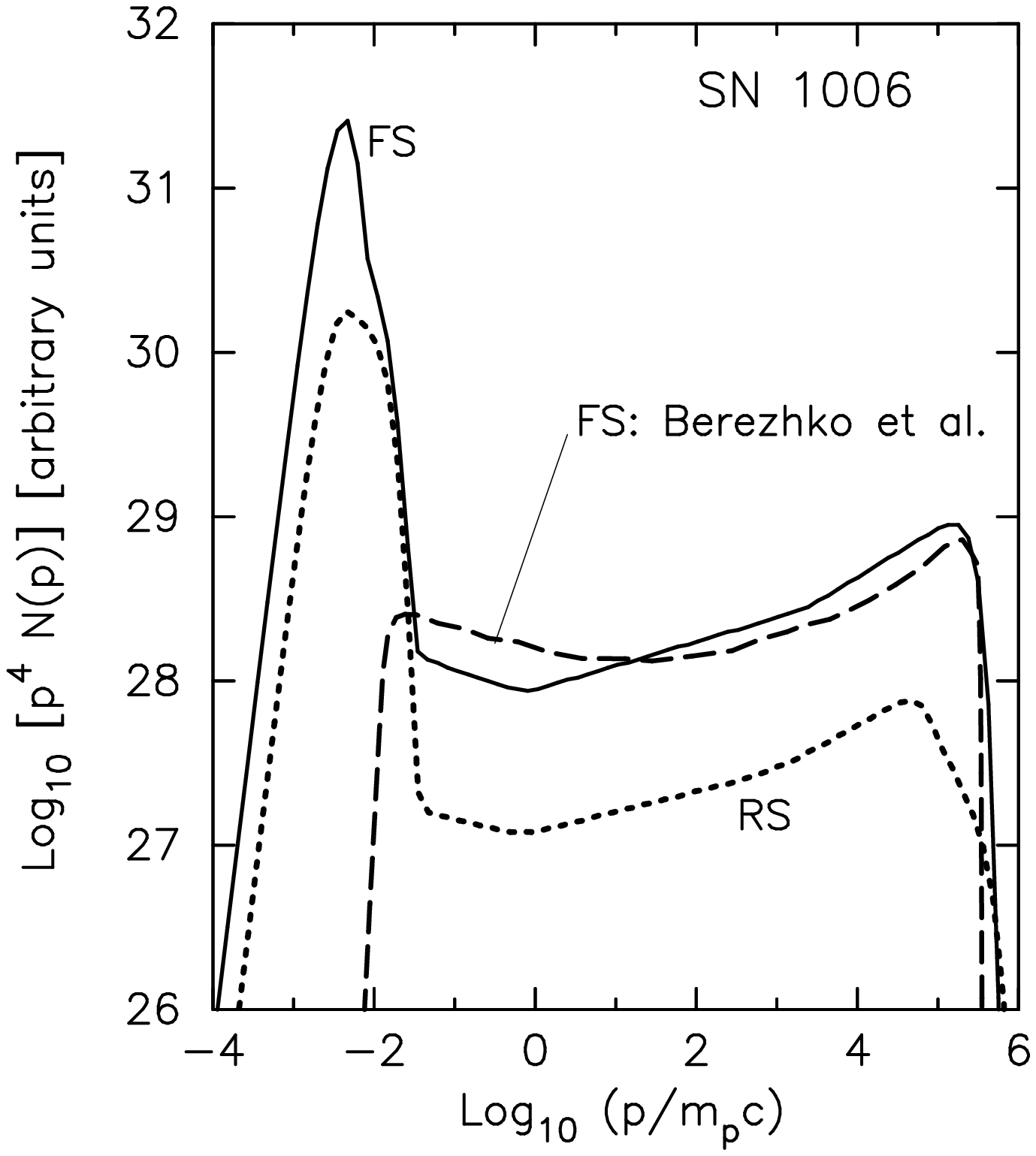} 
\caption{The dashed curve is the inferred FS proton distribution of
  SN1006 as obtained by \citet{BKV2002} at $\tSNR=1000$ yr. Using
  their input parameters for $\EnSN$, $\Mej$, etc., we have matched
  this result (solid curve) with $\etainj=1.2\xx{-4}$, $\alpha=4$, and
  $\fRad = 0.1$ (see equation~\ref{eq:expo} and the discussion
  following it). The
  dotted curve is the corresponding proton distribution from shocked
  ejecta (RS) from our model.
\label{fig:sn1006}
}
\end{figure}

Our CR-hydro model is complementary to that developed by Berezhko and
co-workers in that we use a more approximate model of the NL
acceleration, but have a more complete model of the hydrodynamics
including the reverse shock. In particular, we do not follow the
particles through the acceleration process where previously
accelerated particles can continue to be re-accelerated at later
times. Instead, we assume that all accelerated particles originate as
thermal particles in the upstream flow, are accelerated as the shock
overtakes them, and then remain in the shell of material where they
were accelerated, suffering adiabatic losses as the remnant ages.
The re-acceleration of high energy particles with long diffusion
lengths can be important as it adds to the maximum energy particles
obtain \citep[e.g.][]{BEK96}, but this effect becomes less important
as time goes on and new unshocked particles are overtaken and
accelerated.

In Fig.~\ref{fig:sn1006} we compare our results with those of
\citet{BKV2002} for the proton distribution at $\tSNR=1000$ yr
they predict for SN1006. We have adjusted the normalization of
the FS to obtain a good fit, but otherwise have used the
environmental parameters given in Berezhko et al.
While both our models require an arbitrary
injection parameter defined as the fraction of thermal particles
overtaken by the shock that become superthermal, the actual
implementation of this parameter may be different in the two models. To
fit SN 1006, Berezhko et al. use $\eta =
2\xx{-4}$; we have used $\etainj=1.2\xx{-4}$ to match their
results. 
A distinct advantage of Berezhko's model is that the maximum CR
energy, $\Emax$, and the shape of $f(p)$ near $\Emax$ are determined
self-consistently. In our CR-hydro model, we parameterize these with
$\alpha$ in equation~(\ref{eq:expo}) for the shape and with the
fraction of shock radius, $\fRad$, set equal to the maximum particle
diffusion length to determine $\Emax$. To match the results given by
\citet{BKV2002}, we use $\alpha=4$ and $\fRad=0.1$.
While fine tuning of our parameters could provide a more exact match
than shown in Fig.~\ref{fig:sn1006}, the particular values are
unimportant compared to the fact that both models show a similar level
of injection efficiency and shock modification.
Considering the fundamental differences in the two models, the fit of
superthermal particles accelerated by the FS shown in
Fig.~\ref{fig:sn1006} is impressive. The concave shape, accentuated
by plotting $p^4 N(p)$, is very similar, although the piecewise nature
of our approximation is apparent.

\begin{figure}              % Figure 11
   \centering
   \includegraphics[width=8cm]{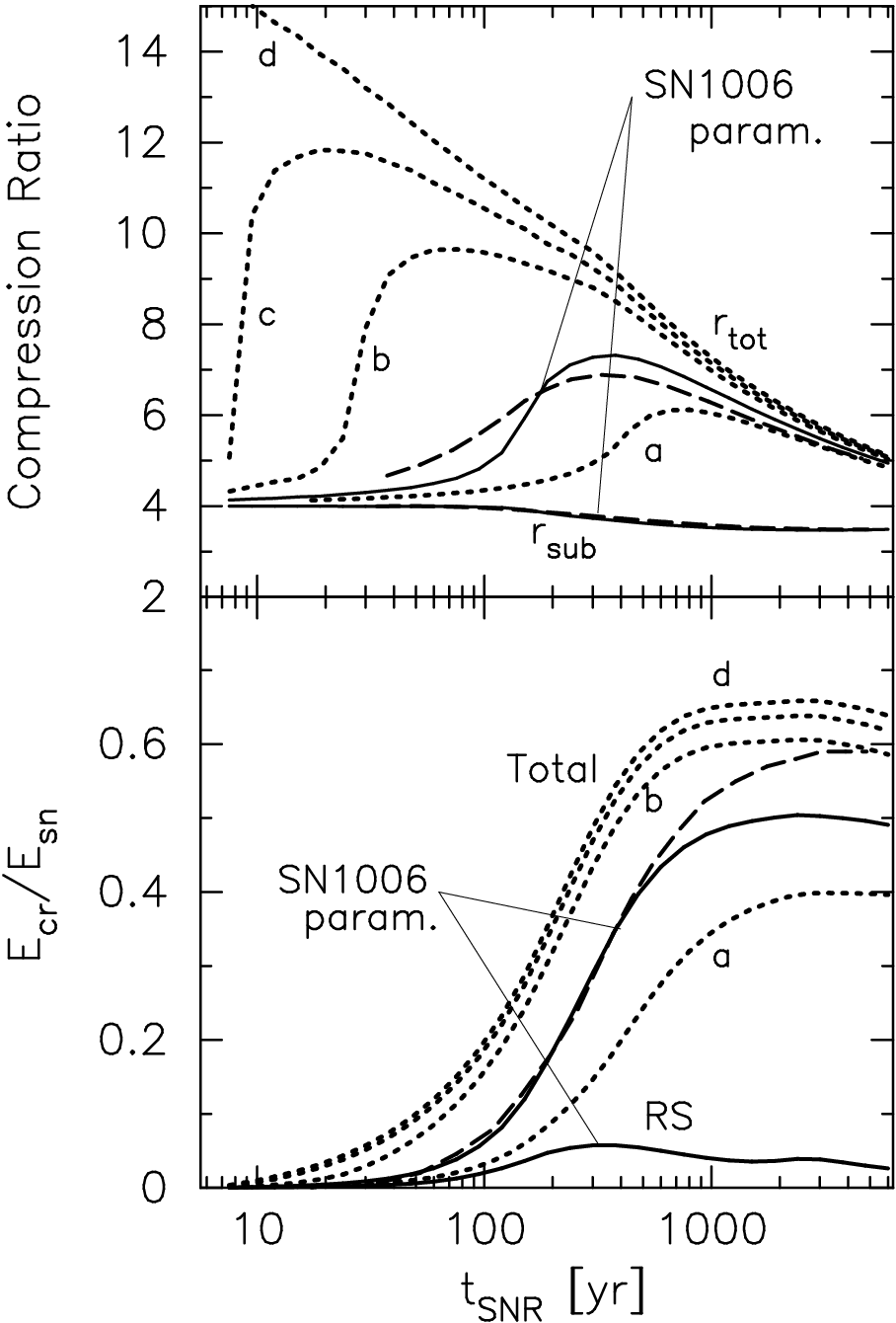} 
\caption{Forward shock compression ratios (top panel) and fraction
  of $\EnSN$ going into relativistic particles, $\EnCR/\EnSN$, (bottom
  panel) for SNRs with various $\etainj$'s versus $\tSNR$. The solid
  curves are for SN1006 parameters ($\etainj=1.2\xx{-4}$, $B_0=20$ \muG,
  $\EnSN=3\xx{51}$ erg, $\Mej=1.4 \Msun$, $\npz =0.3$ \pcc, and
  $\fHe=0.1$) and show $\Rtot$ and $\Rsub$ in the top panel and the
  total $\EnCR/\EnSN$ and that portion from the RS in the bottom
  panel. The dashed curves are the corresponding SN1006 results from
  \citet{BKV2002}. The dotted curves are from the CR-hydro model with
  (a) $\etainj=8\xx{-5}$, (b) $\etainj=2.4\xx{-4}$, (c)
  $\etainj=4\xx{-4}$, and (d) $\etainj=10^{-3}$.
\label{fig:sn1006_comp}
}
\end{figure}

Equally important, the FS compression ratios are very similar, as
shown in the top panel of Fig.~\ref{fig:sn1006_comp} where the total
and subshock compression ratios are compared for this SN1006 model
(CR-hydro: solid curves; Berezhko et al.: dashed curves). As we
discussed earlier, unmodified strong shocks can exist when $\etainj$
is relatively low and this is just what is predicted in both our
models for $\tSNR \la 200$ yr in SN 1006. As the FS slows, the
acceleration becomes more efficient, $\Rtot$ increases, and $\Rsub$
starts to drop below 4 at $\tSNR \ga 200$ yr, indicating that the
shocked proton temperature will be less than predicted in the TP case.
The dotted curves are CR-hydro results with various $\etainj$'s and
show (curves {\it b} and {\it c}) that the transition between
unmodified and modified shocks can be extremely abrupt. For large
enough $\etainj$ (curve {\it d}) unmodified solutions do not occur and
$\Rtot$ is large from the start.

We take the excellent correspondence achieved with Berezhko's
calculation as evidence that our CR-hydro model, with its algebraic NL
particle acceleration approximation, is accurate.

As is clear in Fig.~\ref{fig:sn1006}, \citet{BKV2002} do not treat
thermal particles explicitly although, in principle, they could
convert the thermal pressure and density into a Maxwellian
distribution as we do. They also ignore the reverse shock, an
approximation that is well justified if only continuum emission from
relativistic particles is considered at times $\tSNR \gg \tch$.
In order to predict thermal X-ray emission in young SNRs, however,
thermal electrons must be modeled and particle heating and
acceleration at the forward and reverse shocks must be
self-consistently determined.
We do not show electron spectra here, but our NL acceleration model,
with additional parameters, can produce these along with the
broad-band continuum emission from radio to TeV $\gamma$-rays produced
by them \citep[e.g.,][]{ESG2001}.  The most important advantage our
CR-hydro model has, however, is the ability to model the RS and the
potential for including X-ray thermal emission from the shock-heated
ejecta calculated using a non-equilibrium ionization calculation
\citep[e.g.,][]{DEB2000}.

In the bottom panel of Fig.~\ref{fig:sn1006_comp} we show the
fraction of explosion energy in relativistic particles, $\EnCR/\EnSN$,
as a function of $\tSNR$. As in the top panel, the solid curves are
from our CR-hydro model and the dashed curve (FS only) is from
\citet{BKV2002}. The total $\EnCR/\EnSN$ is in good agreement at early
times but diverges somewhat at later times, a possible indication that
adiabatic losses are treated differently in the two models. Most
importantly, both models show that approximately 50\% of the explosion
energy is placed into cosmic rays during the SNR
lifetime.\footnote{\citet{BKV2002} make the important point that
injection may vary over the surface of the SNR and be significantly
less where the magnetic field is highly oblique. They estimate that to
supply the galactic cosmic-ray population the overall efficiency need
only be $\sim 20$\% of the maximum values shown in
Fig.~\ref{fig:sn1006_comp}.  } The dotted curves in the bottom panel
show that $\EnCR/\EnSN$ at $\tSNR = 6000$ yr remains in a fairly
narrow range ($\sim 40$ to 65\%) for $\etainj$ ranging from $8\xx{-5}$
(a), to $10^{-3}$ (d).

\section{CONCLUSIONS}

The efficient production of cosmic rays by shocks in SNRs lowers the
pressure to energy density ratio in the post-shock gas causing
dramatic changes in the thermal properties of the shocked gas and the
SNR evolution.
We have presented a CR-hydro model that combines a 1-D hydrodynamic
simulation of a SNR, including the forward and reverse shocks, with
particle acceleration. We explicitly include the effects of
particle acceleration on the shock heated ejecta, a critical step in
determining how X-ray thermal emission from the hot ejecta is modified
by particle acceleration.

\newlistroman

In accord with previous results
\citep[e.g.,][]{Dorfi90,BEK96,BKV2002}, we find that SNRs can easily
transfer $\sim 50$\% of the explosion energy into relativistic
particles (bottom panel of Fig.~\ref{fig:sn1006_comp}). Compared to
the situation where acceleration is absent or inefficient, this
transfer results in lower FS speeds, larger overall compression
ratios, cooler post-shock temperatures, and a smaller and denser
interaction region between the forward and reverse shocks.  At young
ages, compression ratios $\Rtot >6$ (top panel of
Fig.~\ref{fig:sn1006_comp}) and proton temperatures less than
$\la 1/10$ the TP value are predicted (Fig.~\ref{fig:temp}) for
the particular set of SN and ISM parameters used here to model SN
1006.
For some injection efficiencies, an abrupt transition is predicted to
occur at early times ($\tSNR < \tch$) between high Mach number,
unmodified shocks with $\Rtot \sim 4$ and strongly modified shocks
with $\Rtot \gg 4$ (top panel of Fig.~\ref{fig:sn1006_comp}),
causing steep spatial gradients in temperature
(Fig.~\ref{fig:TP_NL_stack}).
We believe the changes in shock speed, density, temperature, and
spatial extent and profile are large enough to \listromanDE provide
diagnostics for X-ray observations of young SNRs sufficient to place
meaningful constraints on the acceleration efficiency, and
\listromanDE to importantly modify the inferred values of $\EnSN$,
$\Mej$, and ambient ISM density
once a full thermal X-ray emission model is combined with the CR-hydro
model.

Despite the strong NL effects expected for young SNRs, the CR spectrum
integrated over the age of a remnant (Fig.~\ref{fig:cr_spec}) should
have a spectrum not much flatter than $N(E) \propto E^{-2}$ at
relativistic energies.

Our one-dimensional CR-hydro model uses a computationally fast,
approximate calculation of particle acceleration which is coupled to
the hydrodynamics by modifying the effective ratio of specific
heats. We have verified the accuracy of this approach for parameters
applicable to SN1006 by direct comparison with the more physically
complete model of acceleration of \citet{BKV2002}
(Figs.~\ref{fig:sn1006} and \ref{fig:sn1006_comp}), where the
time-dependent CR transport equations are solved self-consistently
with the gas-dynamic equations.  The main advantage of our model lies
in the fact that we can model acceleration at the forward and reverse
shocks during all stages of the SNR evolution.
In a preliminary work we modeled thermal X-ray emission from Kepler's
SNR with a two-fluid, self-similar description of the SNR
hydrodynamics coupled to the same calculation of particle acceleration
\citep[][]{DEB2000}.  We have demonstrated (Fig.~\ref{fig:SS_comp})
that the two models give essentially identical results when
self-similar conditions apply.  Our next step will be to include this
non-equilibrium calculation of thermal X-ray emission, plus broad-band
continuum emission from \brem, \synch, \IC, and \pion\
\citep[][]{BaringEtal99}, in the CR-hydro model.

\begin{acknowledgements}
The authors would like to thank J. Blondin for
providing his hydrodynamic simulation {\it VH-1} and for help
modifying it to include particle acceleration.
This work was supported, in part, by a NSF-CNRS  grant (NSF INT-0128883)
and by a NASA ATP grant (ATP02-0042-0006).
\end{acknowledgements}

\newcommand\itt{ }
\newcommand\bff{ }
\newcommand{\aaDE}[3]{ 19#1, A\&A, #2, #3}
\newcommand{\aatwoDE}[3]{ 20#1, A\&A, #2, #3}
\newcommand{\aasupDE}[3]{ 19#1, {\itt A\&AS,} {\bff #2}, #3}
\newcommand{\ajDE}[3]{ 19#1, {\itt AJ,} {\bff #2}, #3}
\newcommand{\anngeophysDE}[3]{ 19#1, {\itt Ann. Geophys.,} {\bff #2}, #3}
\newcommand{\anngeophysicDE}[3]{ 19#1, {\itt Ann. Geophysicae,} {\bff #2}, #3}
\newcommand{\annrevDE}[3]{ 19#1, {\itt Ann. Rev. Astr. Ap.,} {\bff #2}, #3}
\newcommand{\apjDE}[3]{ 19#1, {\itt ApJ,} {\bff #2}, #3}
\newcommand{\apjtwoDE}[3]{ 20#1, {\itt ApJ,} {\bff #2}, #3}
\newcommand{\apjletDE}[3]{ 19#1, {\itt ApJ,} {\bff  #2}, #3}
\newcommand{\apjlettwoDE}[3]{ 20#1, {\itt ApJ,} {\bff  #2}, #3}
\newcommand{\apjpress}{{\itt ApJ,} in press}
\newcommand{\apjletpress}{{\itt ApJ(Letts),} in press}
\newcommand{\apjsDE}[3]{ 19#1, {\itt ApJS,} {\bff #2}, #3}
\newcommand{\apjsubDE}[1]{ 19#1, {\itt ApJ}, submitted.}
\newcommand{\apjsubtwoDE}[1]{ 20#1, {\itt ApJ}, submitted.}
\newcommand{\appDE}[3]{ 19#1, {\itt Astropart. Phys.,} {\bff #2}, #3}
\newcommand{\apptwoDE}[3]{ 20#1, {\itt Astropart. Phys.,} {\bff #2}, #3}
\newcommand{\araaDE}[3]{ 19#1, {\itt ARA\&A,} {\bff #2},
   #3}
\newcommand{\assDE}[3]{ 19#1, {\itt Astr. Sp. Sci.,} {\bff #2}, #3}
\newcommand{\grlDE}[3]{ 19#1, {\itt G.R.L., } {\bff #2}, #3} 
\newcommand{\icrcplovdiv}[2]{ 1977, in {\itt Proc. 15th ICRC(Plovdiv)},
   {\bff #1}, #2}
\newcommand{\icrcsaltlake}[2]{ 1999, {\itt Proc. 26th Int. Cosmic Ray Conf.
    (Salt Lake City),} {\bff #1}, #2}
\newcommand{\icrcsaltlakepress}[2]{ 19#1, {\itt Proc. 26th Int. Cosmic Ray Conf.
    (Salt Lake City),} paper #2}
\newcommand{\icrchamburg}[2]{ 2001, {\itt Proc. 27th Int. Cosmic Ray Conf.
    (Hamburg),} {\bff #1}, #2}
\newcommand{\JETPDE}[3]{ 19#1, {\itt JETP, } {\bff #2}, #3}
\newcommand{\jgrDE}[3]{ 19#1, {\itt J.G.R., } {\bff #2}, #3}
\newcommand{\mnrasDE}[3]{ 19#1, {\itt M.N.R.A.S.,} {\bff #2}, #3}
\newcommand{\mnrastwoDE}[3]{ 20#1, {\itt M.N.R.A.S.,} {\bff #2}, #3}
\newcommand{\mnraspress}[1]{ 20#1, {\itt M.N.R.A.S.,} in press}
\newcommand{\natureDE}[3]{ 19#1, {\itt Nature,} {\bff #2}, #3}
\newcommand{\naturetwoDE}[3]{ 20#1, {\itt Nature,} {\bff #2}, #3}
\newcommand{\pfDE}[3]{ 19#1, {\itt Phys. Fluids,} {\bff #2}, #3}
\newcommand{\phyreptsDE}[3]{ 19#1, {\itt Phys. Repts.,} {\bff #2}, #3}
\newcommand{\physrevEDE}[3]{ 19#1, {\itt Phys. Rev. E,} {\bff #2}, #3}
\newcommand{\prlDE}[3]{ 19#1, {\it Phys. Rev. Letts,} {\bf #2}, #3}
\newcommand{\revgeospphyDE}[3]{ 19#1, {\itt Rev. Geophys and Sp. Phys.,}
   {\bff #2}, #3}
\newcommand{\rppDE}[3]{ 19#1, {\itt Rep. Prog. Phys.,} {\bff #2}, #3}
\newcommand{\rpptwoDE}[3]{ 20#1, {\itt Rep. Prog. Phys.,} {\bff #2}, #3}
\newcommand{\ssrDE}[3]{ 19#1, {\itt Space Sci. Rev.,} {\bff #2}, #3}
\newcommand{\ssrtwoDE}[3]{ 20#1, {\itt Space Sci. Rev.,} {\bff #2}, #3}
\newcommand{\scienceDE}[3]{ 19#1, {\itt Science,} {\bff #2}, #3} 
\newcommand{\spDE}[3]{ 19#1, {\itt Solar Phys.,} {\bff #2}, #3} 
\newcommand{\spuDE}[3]{ 19#1, {\itt Sov. Phys. Usp.,} {\bff #2}, #3} 

\end{document}